\documentclass[%
 reprint,
 amsmath,amssymb,
 aps,
pra,
]{revtex4-2}

\usepackage{subfigure}
\usepackage{graphicx}
\usepackage{dcolumn}
\usepackage{bm}
\usepackage{amsmath,amsfonts,amssymb}
\DeclareMathOperator{\sech}{sech}

\begin{document}


\title{Rapid biexciton state preparation in a quantum dot using on-off pulse-sequences}

\author{Dionisis Stefanatos}
\email{dionisis@post.harvard.edu}

\author{Emmanuel Paspalakis}

\affiliation{Materials Science Department, School of Natural Sciences, University of Patras, Patras 26504, Greece}

\date{\today}

\begin{abstract}

We consider the problem of pulsed biexciton preparation in a quantum dot and show that a pulse-sequence with a simple on-off-on modulation can achieve complete preparation of the target state faster than the commonly used constant and hyperbolic secant pulses. The durations of the pulses composing the sequence are obtained from the solution of a transcendental equation. Furthermore, using numerical optimal control, we demonstrate that for a wide range of values of the maximum pulse amplitude, the proposed pulse-sequence prepares the biexciton state in the numerically obtained minimum time, for the specific system under consideration. We finally show with numerical simulations that, even in the presence of dissipation and dephasing, high levels of biexciton state fidelity can be generated in short times. 

\end{abstract}

\maketitle

\section{Introduction}

\label{sec:intro}

The manipulation of exciton and biexciton transitions
in semiconductor quantum dots using laser
pulses is an active research area embodying
various theoretical and experimental investigations \cite{book1}.
Semiconductor quantum dots provide a platform which can be quite useful for quantum information processing, with some important advantages
over other qubit candidates, for example the tailor-made energy
levels, the small decoherence times relative to
the driving times, and the easy integration.

Within this context, a problem which has attracted significant attention is the coherent preparation and in general the coherent population dynamics of a biexciton state, when the quantum dot is initially in its ground state. The coherent preparation of biexciton state by applying laser pulses is very important in the area of quantum information processing, especially for the efficient creation of single photons \cite{Weihs13a} and polarization-entangled photons \cite{Muller14a,Heinze15a,Winik17a,Huber17a,Chen18a} from the exciton-biexciton cascade. Several coherent control methods have been used to achieve this goal, including resonant pulses \cite{Flissikowki04a,Akimov06a,Stufler06a,Machnikowski08a,Paspalakis10a,Brumer13a}, as well as chirped pulses \cite{Hui08a,Axt13a,Amand13a,Kaldewey17a} in order to increase robustness against noise and experimental imperfections. The above methods have succeeded high levels of biexciton fidelity \cite{Kaldewey17a}, even in the presence of phonon-assisted noise processes \cite{Axt14a}. An alternative method, which has also been demonstrated experimentally to reach fidelities over 90\%, actually exploits the phonon-assisted processes to prepare the biexciton state \cite{Ardelt14,Quilter15,Axt15a}. In the current work we concentrate on the basic approach to achieve coherent control of the ground-biexciton transition using a linearly-polarized single laser pulse which drives the exciton-biexciton cascade with a two-photon transition between ground and biexciton states. Within this framework, two frequently used laser pulse shapes are the constant and hyperbolic secant profiles \cite{Paspalakis10a}.

When driving a quantum system from some initial to a final state, it is often desirable to reach the target state in the minimum possible time, at the so called \emph{quantum speed limit} \cite{Mandelstam45,Margolus98,Bhattacharyya83,Anandan90,Deffner13,Deffner17,delCampo13,Shanahan18,Okuyama18,Giovannetti03}. The reason behind this is to reduce the undesirable interactions with the environment, which lead to dissipation and dephasing. Several analytical and numerical optimal control methods \cite{Bryson} have been used to find controls which can drive the quantum systems under consideration at the speed limit \cite{Khaneja01,Boscain06,Carlini06,Caneva09,Bason12,Poggi13,Stefanatos17,Hegerfeldt13,Brouzos15,Hirose18,Fischer19,Lucarelli18,Stefanatos20,Kirchhoff18,Zeng18}. Knowing the quantum speed limit of a system is also important when applying the alternative driving method of \emph{Shortcuts to Adiabaticity (STA)} \cite{Odelin19}, which aims to achieve the same outcome as a slow quantum adiabatic process but in significantly shorter times, since there is a trade-off between speed and energetic cost of implementing STA.

In the present work we show that a simple on-off-on pulse-sequence, with pulse durations obtained from the solution of a transcendental equation, can achieve complete preparation of the biexciton state faster than the commonly used constant and hyperbolic secant pulses. Moreover, using numerical optimal control, we demonstrate that for a wide range of values of the maximum pulse amplitude, the proposed pulse-sequence prepares the biexciton state in the numerically calculated minimum time. This coincidence provides strong evidence that the suggested pulse-sequence drives the specific quantum system, with the control bounded by the maximum pulse amplitude, at its speed limit. We also show with numerical simulations that, even in the presence of dissipation and dephasing, high levels of biexciton state fidelity can be generated in short times. Finally, we point out that numerical optimization can be used to find the controls which maximize the transfer fidelity obtained within a fixed duration, even in the presence of more complicated noise mechanisms like for example phonon-assisted processes.  

The paper is organized as follows. In sections \ref{sec:theory} and \ref{sec:pulse1} we provide the basic theory for biexciton preparation in a quantum dot using a single linearly-polarized pulse and briefly discuss the constant and hyperbolic secant pulses. In Section \ref{sec:pulse2}, which is the main contribution of the present work, we study biexciton state preparation using a on-off-on pulse-sequence and designate the connection with numerical optimal control. In Section \ref{sec:relaxation} we study the effect of dissipation and dephasing, while Section \ref{sec:conclusion} concludes this work.

\section{Pulsed biexciton preparation in a quantum dot}

\label{sec:theory}

If we denote with $|0\rangle, |1\rangle$, and $|2\rangle$ the ground, single-exciton, and biexciton states, respectively, then the Hamiltonian of the biexciton system in the dipole approximation becomes \cite{Machnikowski08a,Paspalakis10a,Hui08a}
\begin{eqnarray}
H_b(t) &=& {E}|1\rangle \langle 1| + (2{ E}+{E_{b}})|2\rangle \langle 2|\nonumber \\
    && -\mu {\cal E_{SQD}}(t)(|0\rangle \langle 1| + |1\rangle \langle 2| + H.c. ),
\label{ham1}
\end{eqnarray}
where $E$ is the single-exciton energy, $E_b$ is
the biexciton energy shift (ground state energy is set to zero), $\mu$ is the dipole moment of the semiconductor quantum dot corresponding to the ground-exciton and exciton-biexciton transitions (assumed for simplicity to be the same for both transitions), and ${\cal E_{SQD}}$ is the external electric field in the semiconductor quantum dot. Note that we have considered a symmetric quantum dot and due to selection rules there is no direct ground-biexciton transition with a single photon. The electric field is expressed as
\begin{equation}
\label{control}
{\cal E_{SQD}}(t) = \frac{\hbar}{\mu}\left[\frac{\Omega(t)}{2}e^{-i\omega t} + H.c. \right],
\end{equation}
where $\Omega(t)$ is the time-dependent Rabi frequency and $\omega$ is the angular frequency of the applied field. $\Omega(t)$ is the control that will be used to generate the biexciton state starting from the ground state. In the following analysis we use a real $\Omega(t)$ since we seek the minimum-time evolution, which is commonly achieved using resonant pulses \cite{Boscain02}, while chirping is usually exploited in order to increase robustness, not speed.

If we use the transformed  probability amplitudes for states $|0\rangle, |1\rangle, |2\rangle$, defined as $\tilde{c}_0=c_0,\tilde{c}_1=c_1e^{i\omega t}, \tilde{c}_2=c_2e^{2i\omega t}$, and perform the rotating wave approximation, we obtain the transformed Hamiltonian
\begin{equation}
\tilde{H}_b(t)=\hbar
\left(
\begin{array}{ccc}
0 & -\frac{\Omega*}{2}  & 0\\
-\frac{\Omega}{2} & \frac{E}{\hbar}-\omega & -\frac{\Omega*}{2} \\
0 & -\frac{\Omega}{2}  & \frac{2E+E_b}{\hbar}-2\omega
\end{array}
\right).
\end{equation}
If, additionally, we set the frequency $\omega$ of the field to satisfy the two-photon resonance condition $\hbar\omega=E+E_b/2$, then we end up with
\begin{equation}
\label{H_trans}
\tilde{H}_b(t)=\hbar
\left(
\begin{array}{ccc}
0 & -\frac{\Omega*}{2}  & 0\\
-\frac{\Omega}{2} & -\frac{E_b}{2\hbar} & -\frac{\Omega*}{2} \\
0 & -\frac{\Omega}{2}  & 0
\end{array}
\right).
\end{equation}
Defining
\begin{subequations}
\label{transformation}
\begin{eqnarray}
A &=& \frac{1}{\sqrt{2}}(\tilde{c}_2+\tilde{c}_0),\\
B &=& \tilde{c}_1,\\
C &=& \frac{1}{\sqrt{2}}(\tilde{c}_2-\tilde{c}_0),
\end{eqnarray}
\end{subequations}
then we easily find that $\dot{C}=0$, while $A, B$ satisfy the two-level system
\begin{equation}
\label{AB_dynamics}
i
\left(
\begin{array}{c}
\dot{A}\\
\dot{B}
\end{array}
\right)
=
\left(
\begin{array}{cc}
0 & -\frac{\Omega}{\sqrt{2}}\\
-\frac{\Omega}{\sqrt{2}} & -\frac{E_b}{2\hbar}
\end{array}
\right)
\left(
\begin{array}{c}
A\\
B
\end{array}
\right).
\end{equation}
This equation can be expressed in compact notation as
\begin{equation}
\label{eq_two_level}
i\dot{\mathbf{a}}=[-\omega_bI+H(t)]\mathbf{a},
\end{equation}
where $\mathbf{a}=(A\quad B)^T$,
\begin{equation}
\label{omegab}
\omega_b=\frac{E_b}{4\hbar},
\end{equation}
$I$ is the $2\times 2$ identity matrix and $H$ is the two-level Hamiltonian
\begin{equation}
\label{H_two_level}
H(t)=
\left(
\begin{array}{cc}
\omega_b & -\frac{\Omega}{\sqrt{2}}\\
-\frac{\Omega}{\sqrt{2}} & -\omega_b
\end{array}
\right)
=-\frac{\Omega(t)}{\sqrt{2}}\sigma_x+\omega_b\sigma_z.
\end{equation}

From the initial conditions $c_0(0)=1, c_1(0)=c_2(0)=0$, which are the same for $\tilde{c}_i$, we get $A(0)=1/\sqrt{2}, B(0)=0, C(0)=-1/\sqrt{2}$. Since $C(t)$ is constant, from Eq. (\ref{transformation}) we see that if the control $\Omega(t)$ is selected such that at the final time $t=T$ we have $A(T)=-1/\sqrt{2}$, then $\tilde{c}_2(T)=-1\Rightarrow|c_2(T)|=1$ and thus complete preparation of the biexciton state is achieved. Note that because of the initial condition, the normalization for the two-level system is $1/\sqrt{2}$ instead of the usual $1$, thus $B(T)=0$. According to the above analysis, the initial and target states for $\mathbf{a}$ are
\begin{equation}
\label{i_f_conditions}
\mathbf{a}(0)=
\left(
\begin{array}{c}
\frac{1}{\sqrt{2}}\\
0
\end{array}
\right),
\quad
\mathbf{a}(T)=-
\left(
\begin{array}{c}
\frac{1}{\sqrt{2}}\\
0
\end{array}
\right),
\quad
\end{equation}
which obviously differ by a $\pi$-phase. If $U$ is the propagator corresponding to Hamiltonian $H$, then the desired phase difference at the final time $t=T$ is accomplished when the total propagator corresponding to Eq. (\ref{eq_two_level}), which includes the effect of the term $-\omega_bI$, satisfies
\begin{equation}
\label{phase_condition}
e^{i\omega_bT}U=
\left(
\begin{array}{cc}
-1 & 0\\
0 & \mbox{indif.}
\end{array}
\right).
\end{equation}
We will use this condition in the following sections.

\section{Constant and hyperbolic secant pulses}

\label{sec:pulse1}

\begin{figure*}[t]
 \centering
		\begin{tabular}{cc}
     	\subfigure[$\ $]{
	            \label{fig:const_pulse}
	            \includegraphics[width=.45\linewidth]{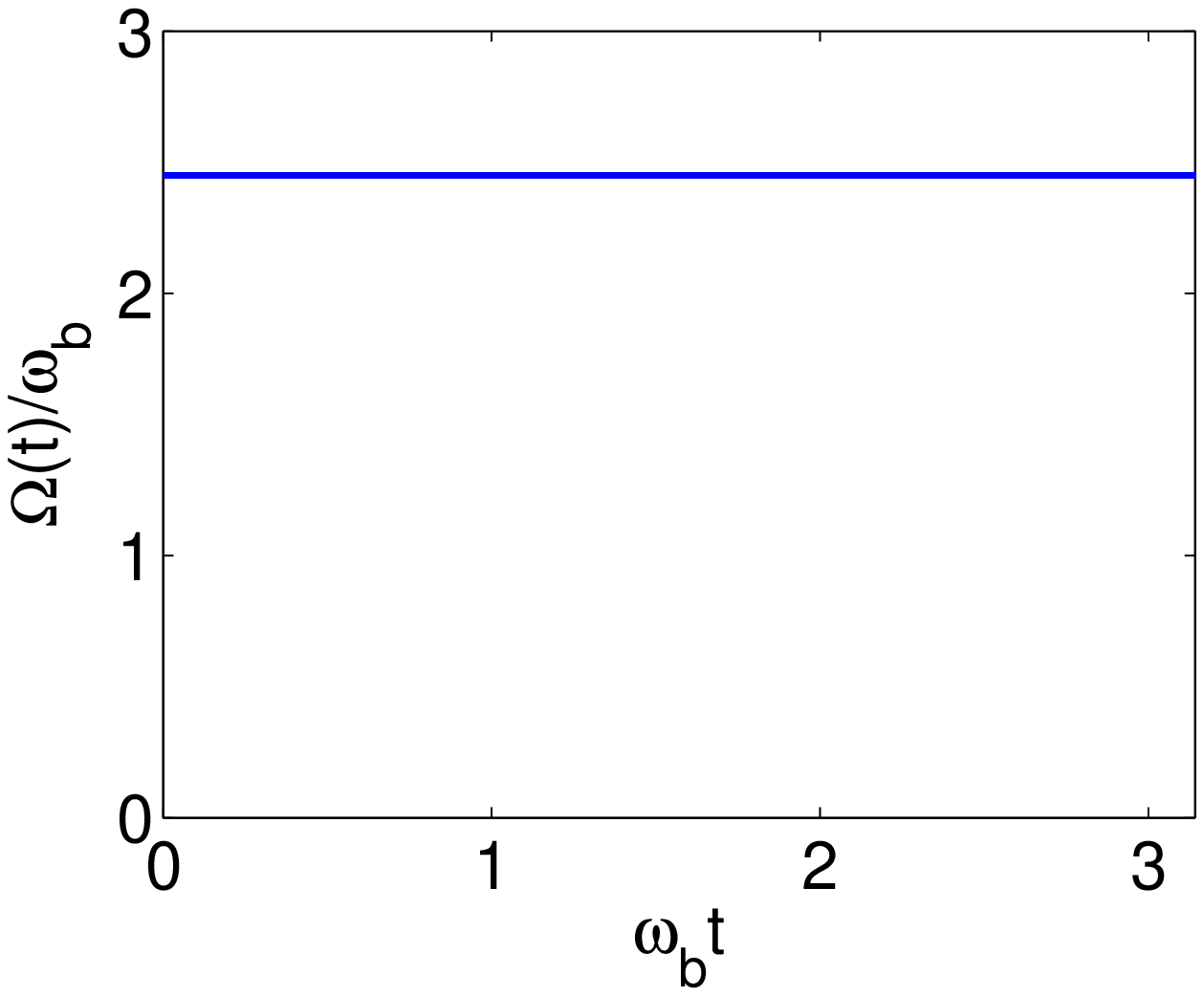}} &
       \subfigure[$\ $]{
	            \label{fig:sech_pulse}
	            \includegraphics[width=.45\linewidth]{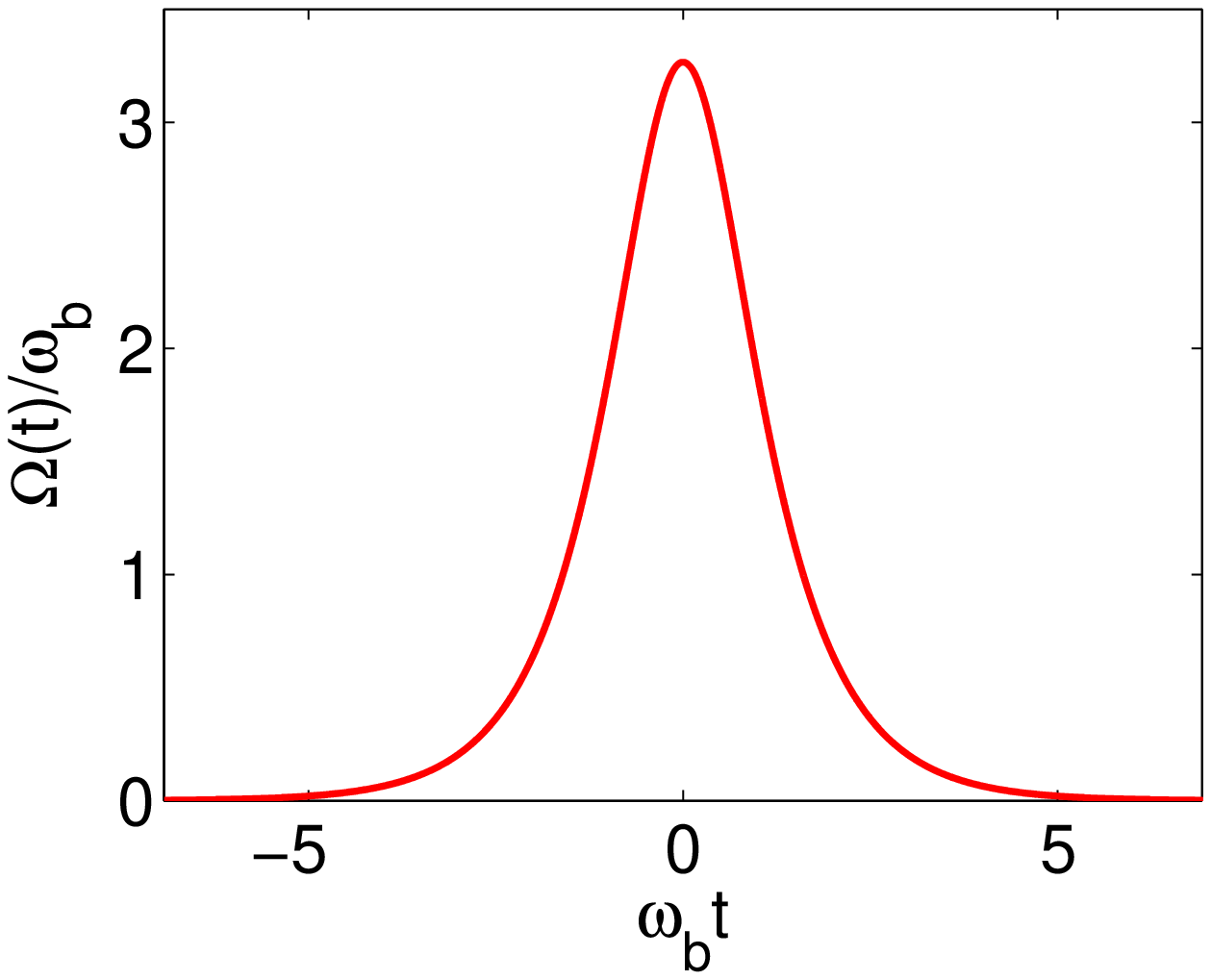}} \\
       \subfigure[$\ $]{
	            \label{fig:const_pop}
	            \includegraphics[width=.45\linewidth]{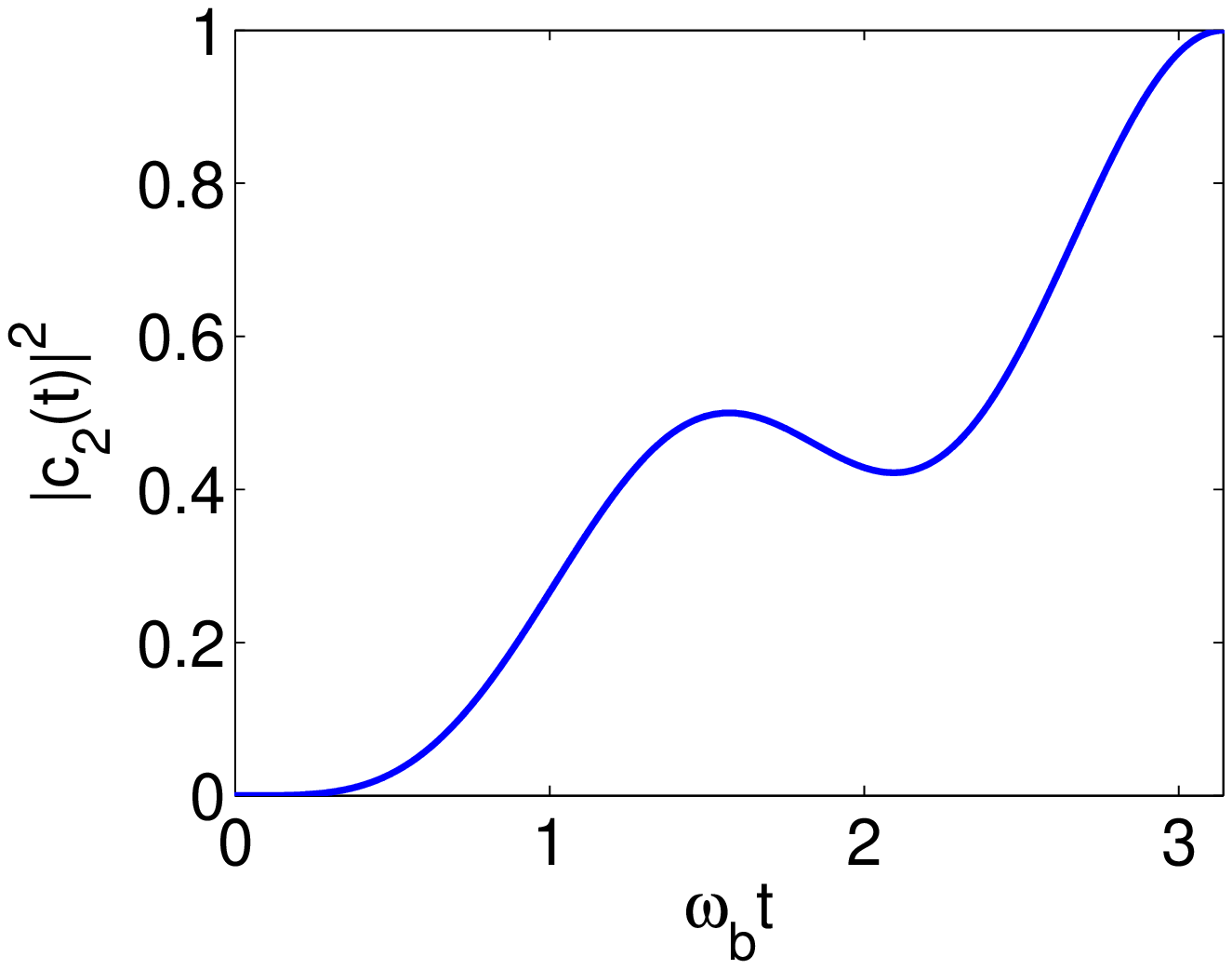}} &
       \subfigure[$\ $]{
	            \label{fig:sech_pop}
	            \includegraphics[width=.45\linewidth]{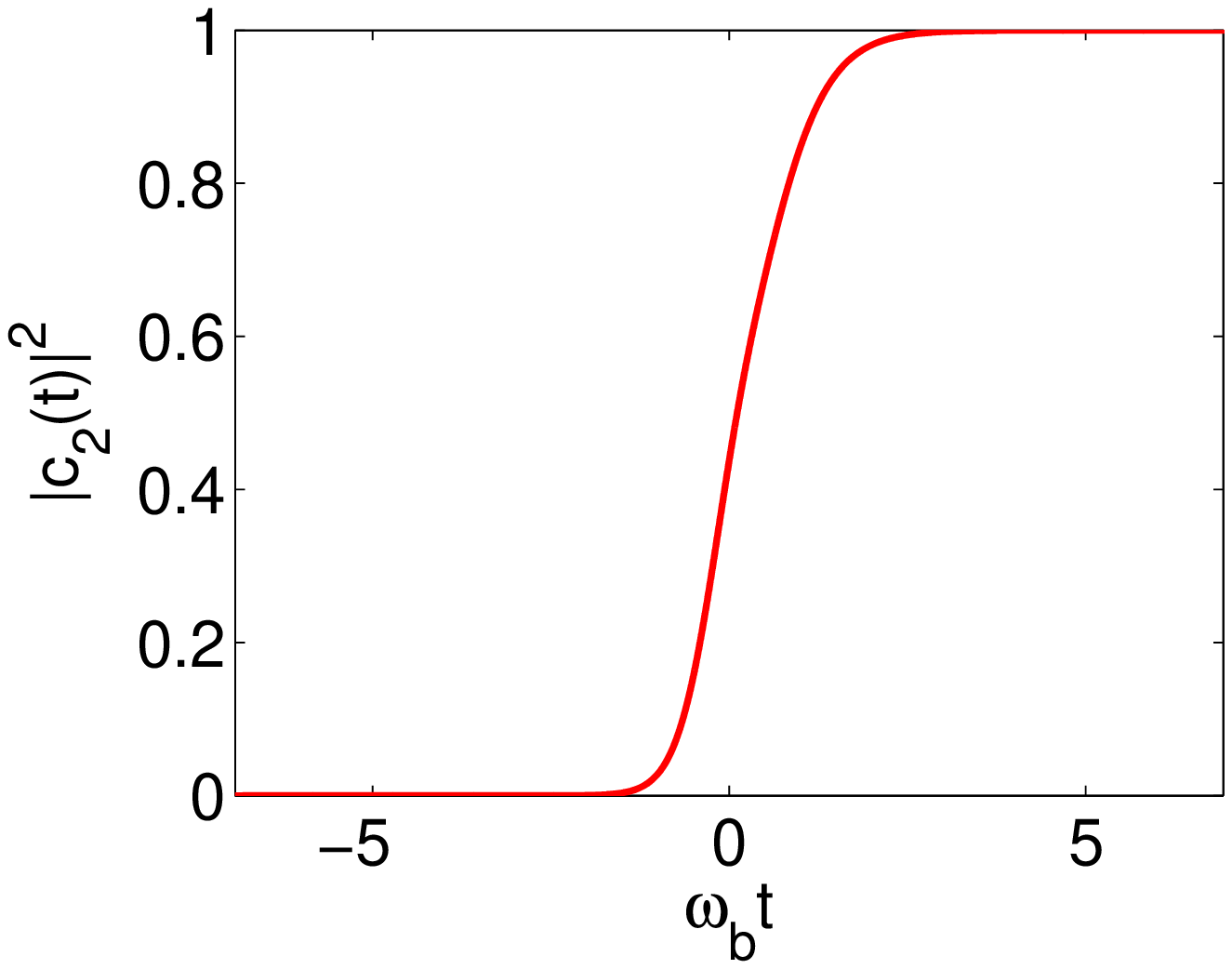}}
		\end{tabular}
\caption{(Color online) (a, c) Shortest-amplitude constant pulse achieving complete biexciton preparation within duration $T=\pi/\omega_b$, and corresponding evolution of population $|c_2(t)|^2$. (b, d) Shortest-width hyperbolic secant pulse achieving complete biexciton preparation, and corresponding evolution of population $|c_2(t)|^2$.}
\label{fig:const_sech}
\end{figure*}

For a constant pulse $\Omega(t)=\Omega_0$ of duration $T$ the corresponding two-level Hamiltonian $H_{on}=-\frac{\Omega_0}{\sqrt{2}}\sigma_x+\omega_b\sigma_z$ is also constant
and the propagator $U$ is
\begin{eqnarray}
\label{U_on}
U&=&U_{on}^T=e^{-iH_{on}T}\nonumber\\
&=&e^{-i\omega T(n_x\sigma_x+n_z\sigma_z)}\nonumber\\
&=& I\cos{\omega T}-i\sin{\omega T}(n_x\sigma_x+n_z\sigma_z)\\
&=&
\left(
\begin{array}{cc}
\cos{\omega T}-in_z\sin{\omega T} & -in_x\sin{\omega T}\\
-in_x\sin{\omega T} & \cos{\omega T}+in_z\sin{\omega T}
\end{array}
\right),\nonumber
\end{eqnarray}
where
\begin{subequations}
\begin{eqnarray}
\label{parameters}
\omega&=&\sqrt{\omega_b^2+\frac{\Omega_0^2}{2}},\\
n_x&=&-\frac{1}{\sqrt{2}}\frac{\Omega_0}{\omega},\\
n_z&=&\frac{\omega_b}{\omega}.
\end{eqnarray}
\end{subequations}
The requirement to be zero the off-diagonal elements in Eq. (\ref{phase_condition}) leads to the condition $\sin{\omega T}=0\Rightarrow\omega T=k\pi$, $k=1,2,\ldots$, thus $\cos{\omega T}=\cos{k\pi}$. Then, the condition for the first diagonal element in Eq. (\ref{phase_condition}) becomes $e^{i\omega_bT}\cos{k\pi}=-1\Rightarrow e^{i\omega_bT}=(-1)^{k-1}=\pm 1$. The shortest duration such that this condition is satisfied is $T=\pi/\omega_b$. Integer $k$ should be even, thus $k=2$ is the smallest possible value, corresponding to the smallest possible amplitude $\Omega_0^{min}$, which satisfies $\omega T=2\pi$. Thus, the shortest duration and the smallest amplitude of a constant pulse which achieves complete biexciton preparation are
\begin{equation}
\label{minT_constant}
T=\frac{\pi}{\omega_b}, \quad \Omega_0^{min}=\omega_b\sqrt{6}\approx 2.45\omega_b.
\end{equation}
In Figs. \ref{fig:const_pulse}, \ref{fig:const_pop} we plot this constant pulse and the corresponding evolution of population $|c_2(t)|^2$, respectively.

Another pulse-shape which has been used for biexciton preparation is hyperbolic secant, $\Omega(t)=\Omega_0\sech{(t/t_p)}$, where $t_p$ characterizes the pulse width. If we apply this pulse in system (\ref{AB_dynamics}) and follow the Rosen-Zener methodology \cite{Rosen32}, using the transformation
\begin{equation}
\label{z}
z(t)=\frac{1}{2}\left[\tanh{\left(\frac{t}{t_p}\right)}+1\right],
\end{equation}
we obtain for $A(z)$ the hypergeometric differential equation
\begin{equation}
\label{hypergeometric}
z(1-z)\frac{d^2A}{dz^2}+[c-(a+b+1)z]\frac{dA}{dz}-abA=0,
\end{equation}
where
\begin{equation}
\label{parameters}
a=-b=\frac{\Omega_0t_p}{\sqrt{2}}, \quad c=\frac{1}{2}-i\omega_bt_p.
\end{equation}
The solution
\begin{equation}
\label{F}
A(z)=\frac{1}{\sqrt{2}}{}_{2}F_1(a,b;c;z),
\end{equation}
where ${}_{2}F_1$ is the hypergeometric function, satisfies the initial condition $A(0)=1/\sqrt{2}$, corresponding to $t\rightarrow-\infty$, see Eq. (\ref{i_f_conditions}). The other initial condition $B(0)=0$, which is translated to $dA/dt|_{t\rightarrow-\infty}=0$ through Eq. (\ref{AB_dynamics}), is automatically satisfied at $z=0$ and finite $dA/dz|_{z=0}$ since
\begin{equation}
\frac{dA}{dt}=\frac{2}{t_p}z(1-z)\frac{dA}{dz}.
\end{equation}
For the final value at $z=1$ ($t\rightarrow\infty$), we find from Gauss' theorem
\begin{eqnarray*}
A(1)&=&\frac{1}{\sqrt{2}}{}_{2}F_1(a,b;c;1)=\frac{1}{\sqrt{2}}\frac{\Gamma(c)\Gamma(c-a-b)}{\Gamma(c-a)\Gamma(c-b)}\\
    &=&\frac{1}{\sqrt{2}}\frac{\Gamma^2\left(\frac{1}{2}-i\omega_bt_p\right)}{\Gamma\left(\frac{1}{2}-\frac{\Omega_0t_p}{\sqrt{2}}-i\omega_bt_p\right)\Gamma\left(\frac{1}{2}+\frac{\Omega_0t_p}{\sqrt{2}}-i\omega_bt_p\right)},
\end{eqnarray*}
where $\Gamma$ is the gamma function. A simplified expression is obtained for
\begin{equation}
\label{sech_amplitude}
\Omega_0=\frac{\sqrt{2}n}{t_p},\quad n=1, 2,\ldots
\end{equation}
using the recurrence relation $\Gamma(z+1)=\Gamma(z)$, where we find
\begin{equation}
\label{final_value}
A(1)=\frac{(-1)^{n}}{\sqrt{2}}\prod^{n-1}_{j=0}\frac{j+\frac{1}{2}+i\omega_b}{j+
\frac{1}{2}-i\omega_b}.
\end{equation}
The pulse duration $t_p$ is obtained as the solution of the final condition
\begin{equation}
\label{final_condition}
A(1)=-\frac{1}{\sqrt{2}}.
\end{equation}
For example, for $n=2$ we find from Eqs. (\ref{final_condition}), (\ref{final_value}), and (\ref{sech_amplitude})
\begin{equation}
\label{n_2}
t_p=\frac{\sqrt{3}}{2}\frac{1}{\omega_b},\quad \Omega_0=4\sqrt{\frac{2}{3}}\omega_b.
\end{equation}
The hyperbolic secant pulse with parameters given in Eq. (\ref{n_2}) is plotted in Fig. \ref{fig:sech_pulse}, while the corresponding evolution of population $|c_2(t)|^2$ is displayed in Fig. \ref{fig:sech_pop}. For larger $n$ the width of the pulse and thus the time for complete transfer to the biexciton state increase.

\section{On-off pulses and numerical optimal control}

\label{sec:pulse2}

\begin{figure*}[t]
 \centering
		\begin{tabular}{cc}
     	
      \subfigure[$\ $]{
	            \label{fig:min_time}
	            \includegraphics[width=.45\linewidth]{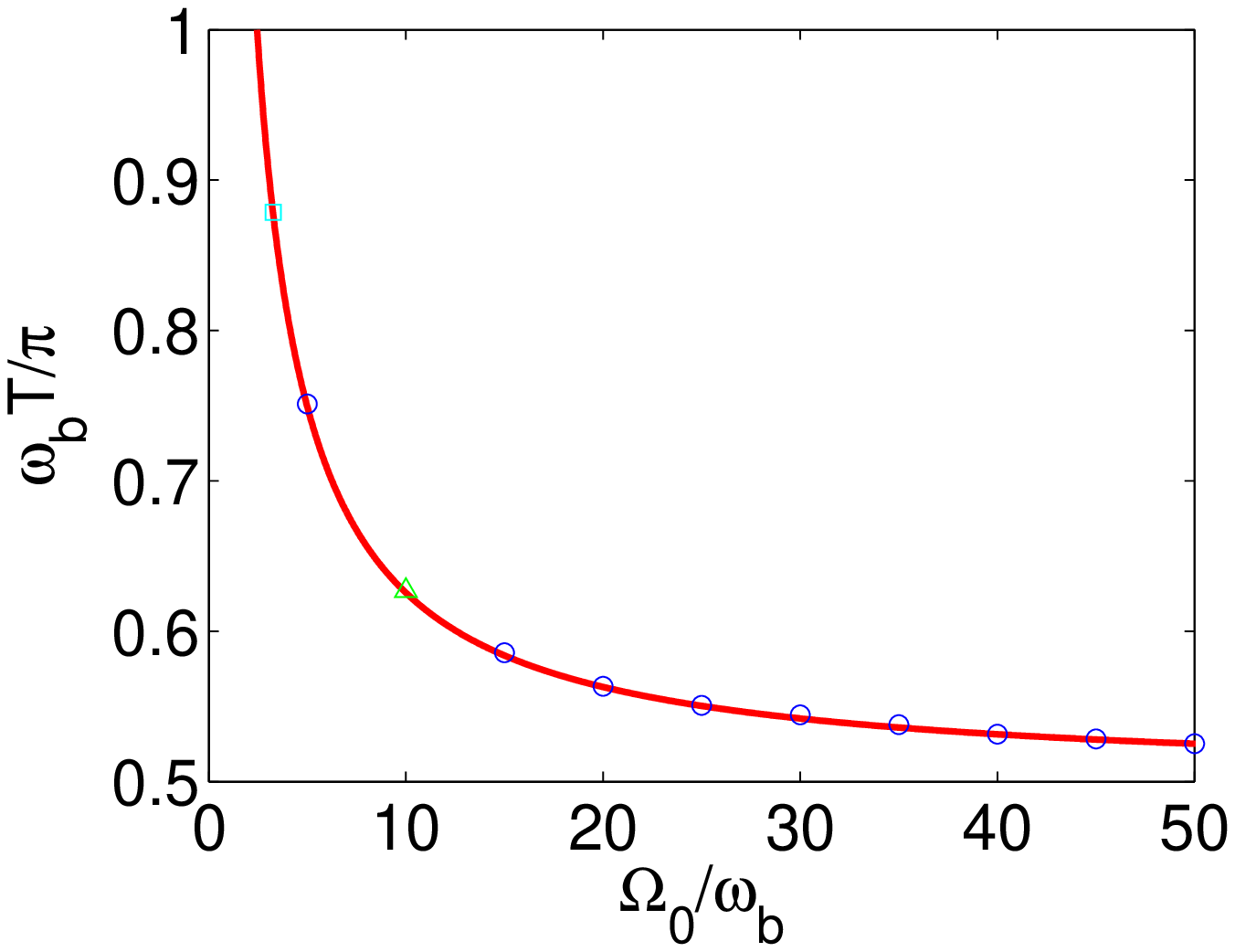}} &
      \subfigure[$\ $]{
	            \label{fig:speed_lim_1}
	            \includegraphics[width=.45\linewidth]{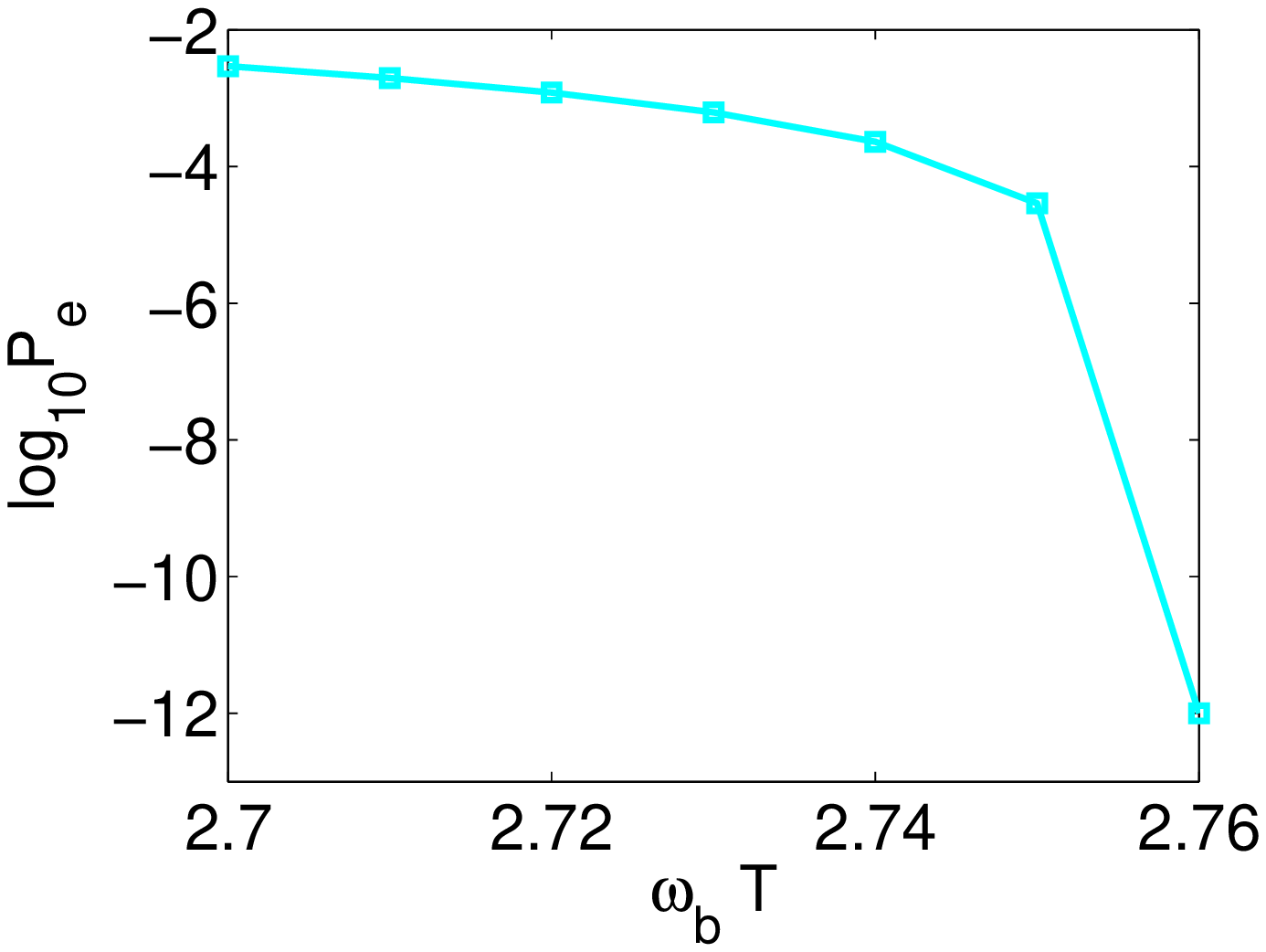}} \\
       \subfigure[$\ $]{
	            \label{fig:speed_lim_2}
	            \includegraphics[width=.45\linewidth]{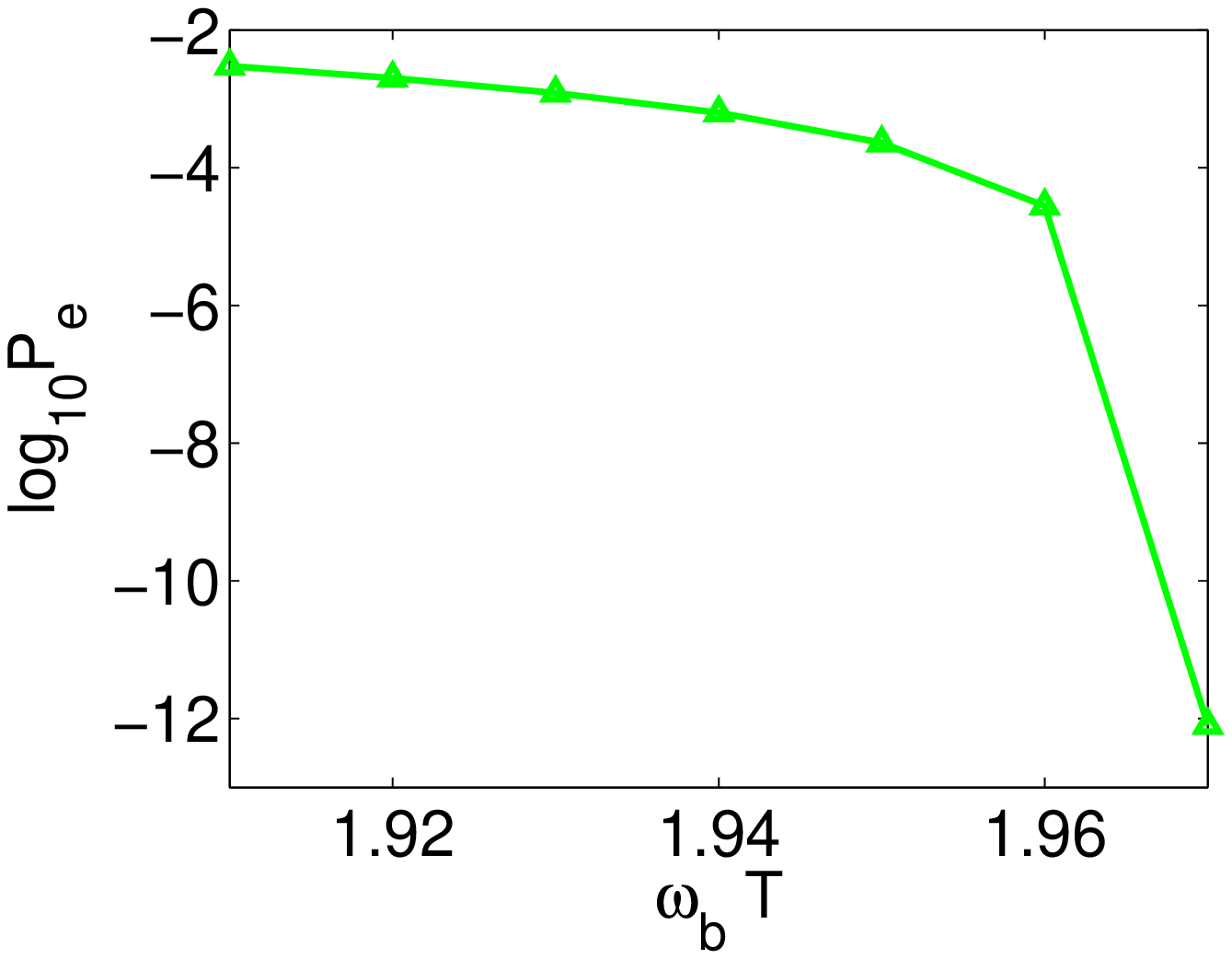}} &
       \subfigure[$\ $]{
	            \label{fig:speed_lim_3}
	            \includegraphics[width=.45\linewidth]{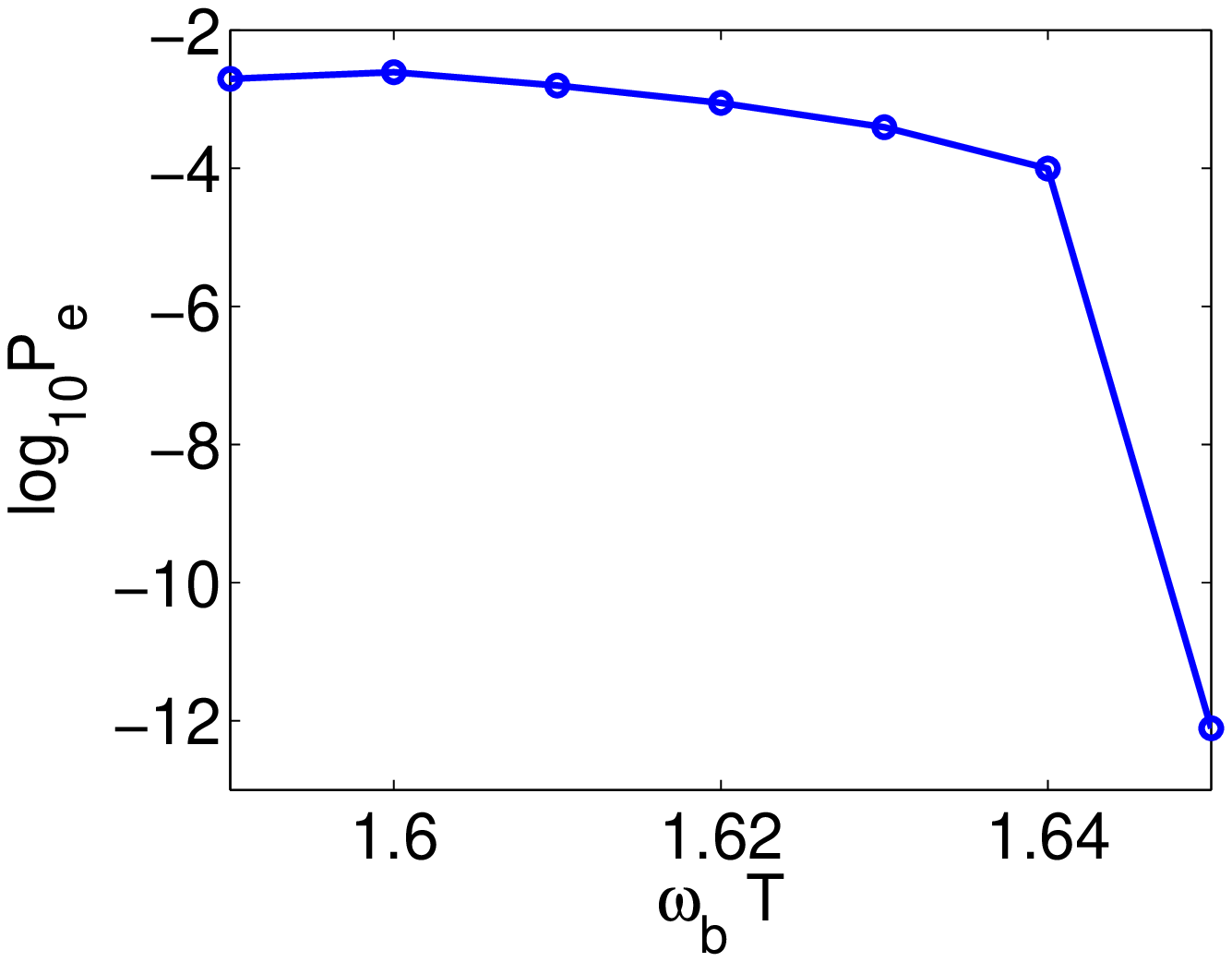}}
		\end{tabular}
\caption{(Color online) (a) Duration of the minimum-time on-off-on pulse-sequence as a function of maximum amplitude $\Omega_0$ (red solid line). For specific values of $\Omega_0$ is also displayed the minimum transfer time obtained with numerical optimization for the specific system under consideration. (b) Numerical derivation of the minimum time for $\Omega_0=4\sqrt{2/3}\omega_b$. As the duration increases from $T=2.75/\omega_b$ to $T=2.76/\omega_b$, the final error $P_e=1-|c_2(T)|^2$ drops substantially, marking the minimum necessary time for biexciton preparation when using the specific maximum amplitude. The corresponding duration $T=2.76/\omega_b$ is then plotted in Fig. \ref{fig:min_time} (cyan square). (c, d) Numerical derivation of the minimum time for $\Omega_0=10/\omega_b$ and $\Omega_0=50/\omega_b$, respectively.}
\label{fig:speed_limit}
\end{figure*}

\begin{figure*}[t]
 \centering
		\begin{tabular}{cc}
     	\subfigure[$\ $]{
	            \label{fig:trans3}
	            \includegraphics[width=.45\linewidth]{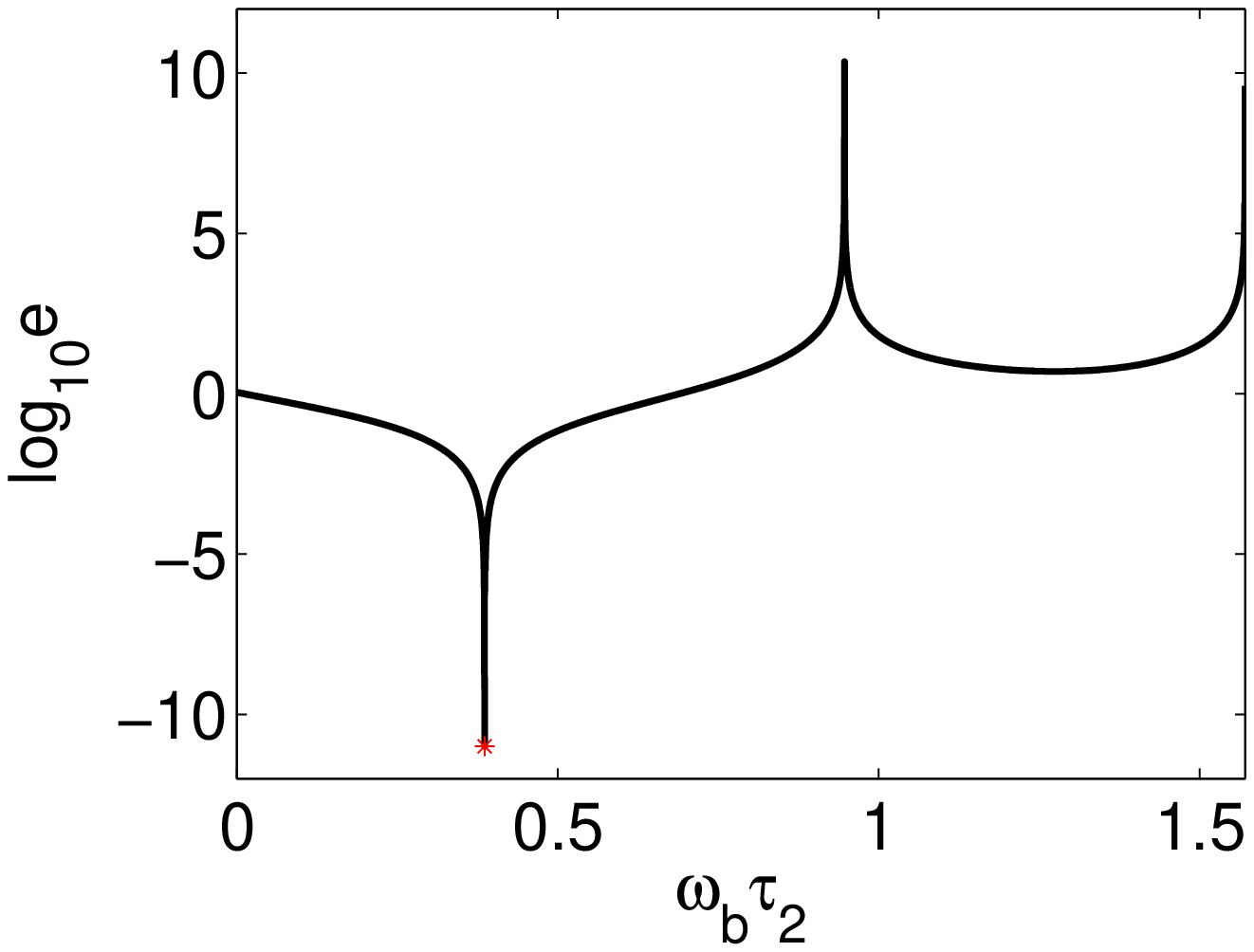}} &
       \subfigure[$\ $]{
	            \label{fig:trans10}
	            \includegraphics[width=.45\linewidth]{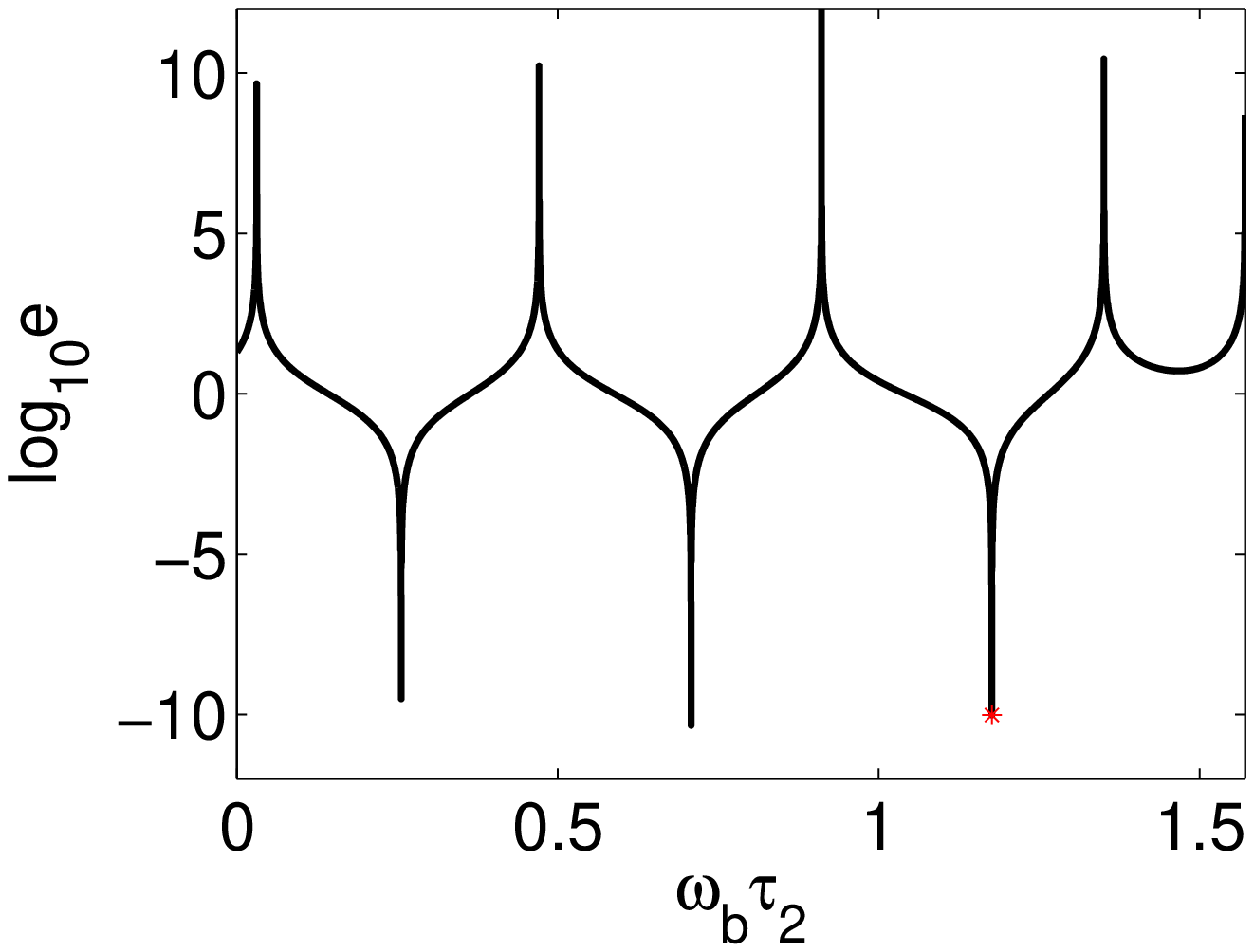}} \\
      \subfigure[$\ $]{
	            \label{fig:control3}
	            \includegraphics[width=.45\linewidth]{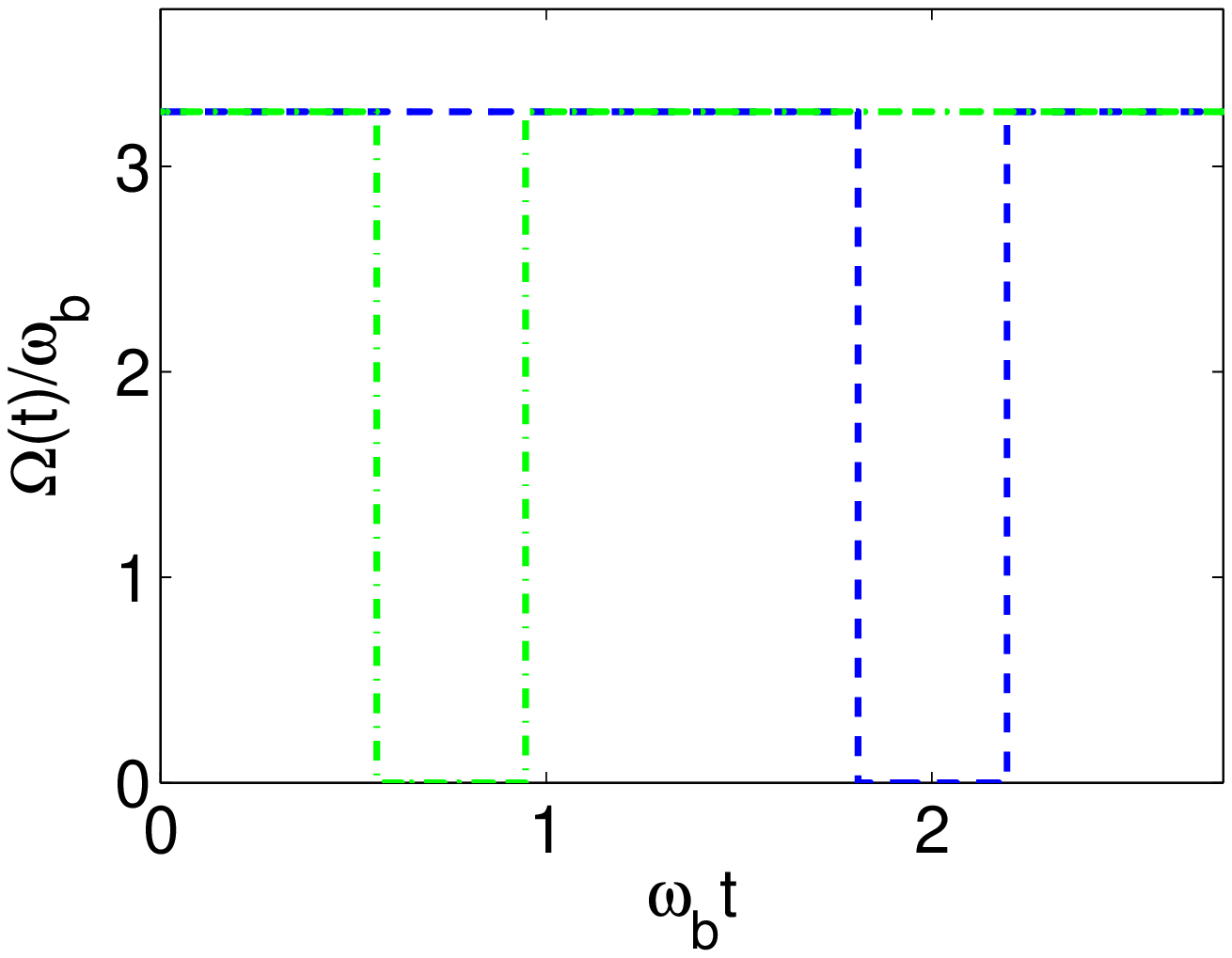}} &
       \subfigure[$\ $]{
	            \label{fig:control10}
	            \includegraphics[width=.45\linewidth]{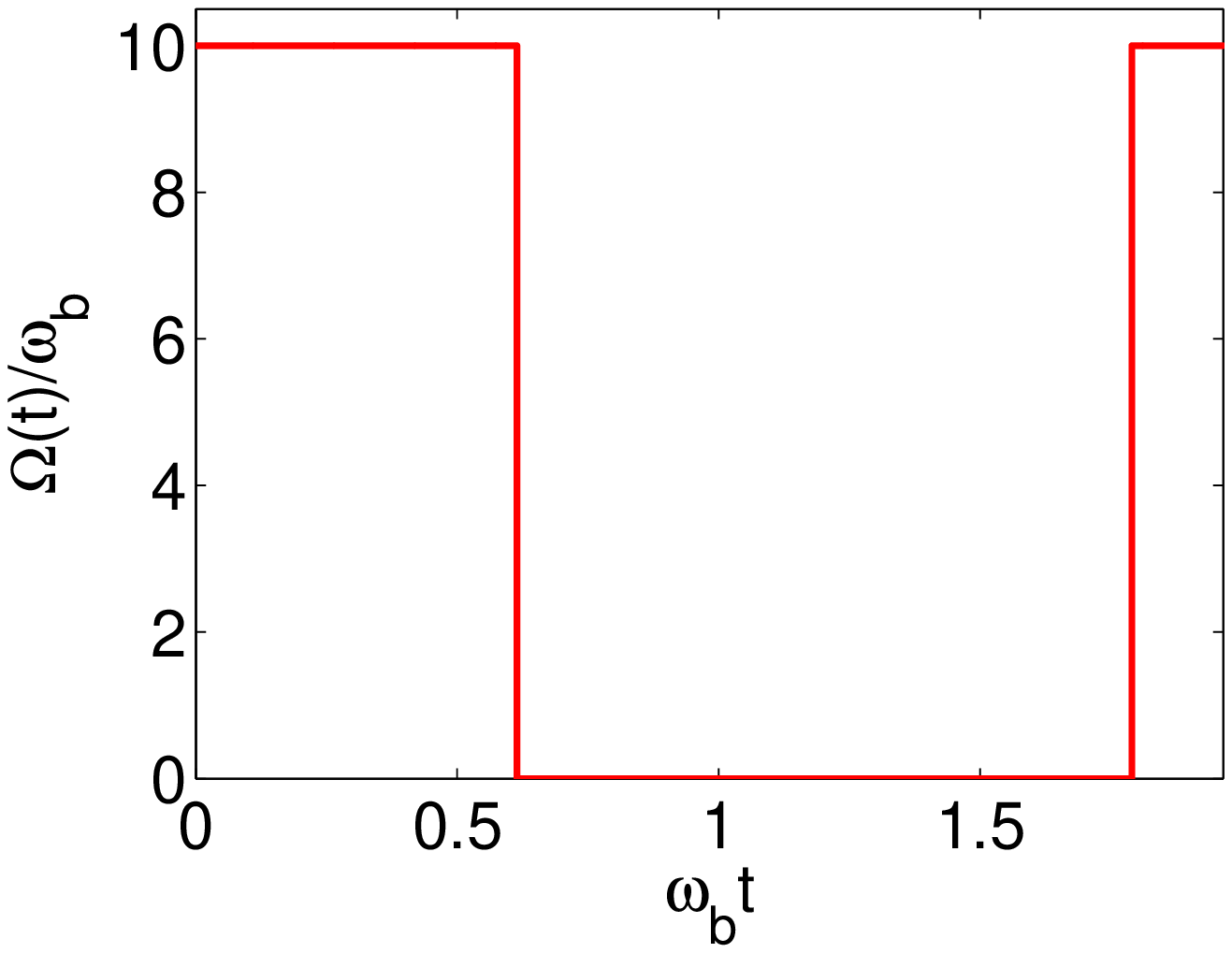}} \\
       \subfigure[$\ $]{
	            \label{fig:population3}
	            \includegraphics[width=.45\linewidth]{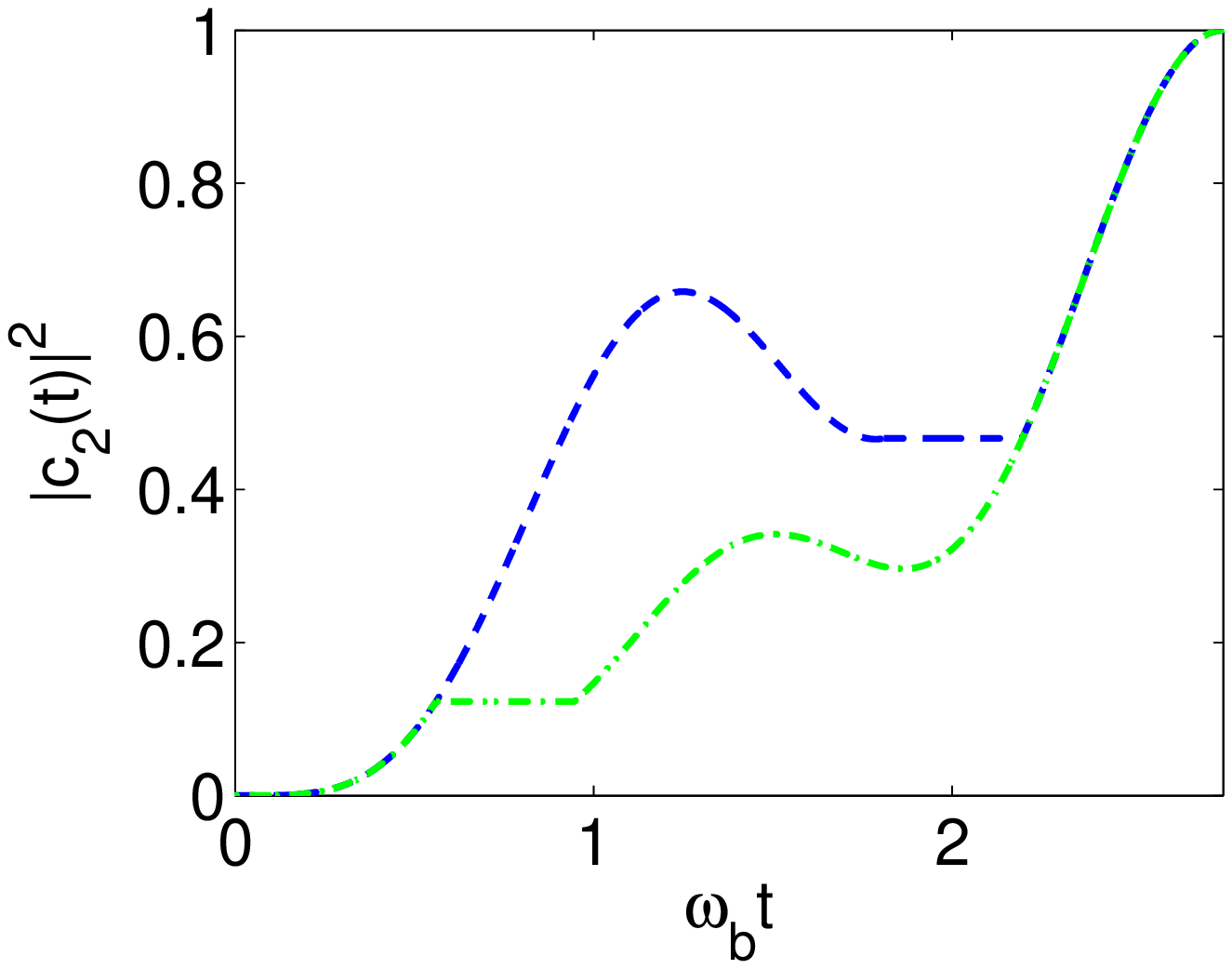}} &
       \subfigure[$\ $]{
	            \label{fig:population10}
	            \includegraphics[width=.45\linewidth]{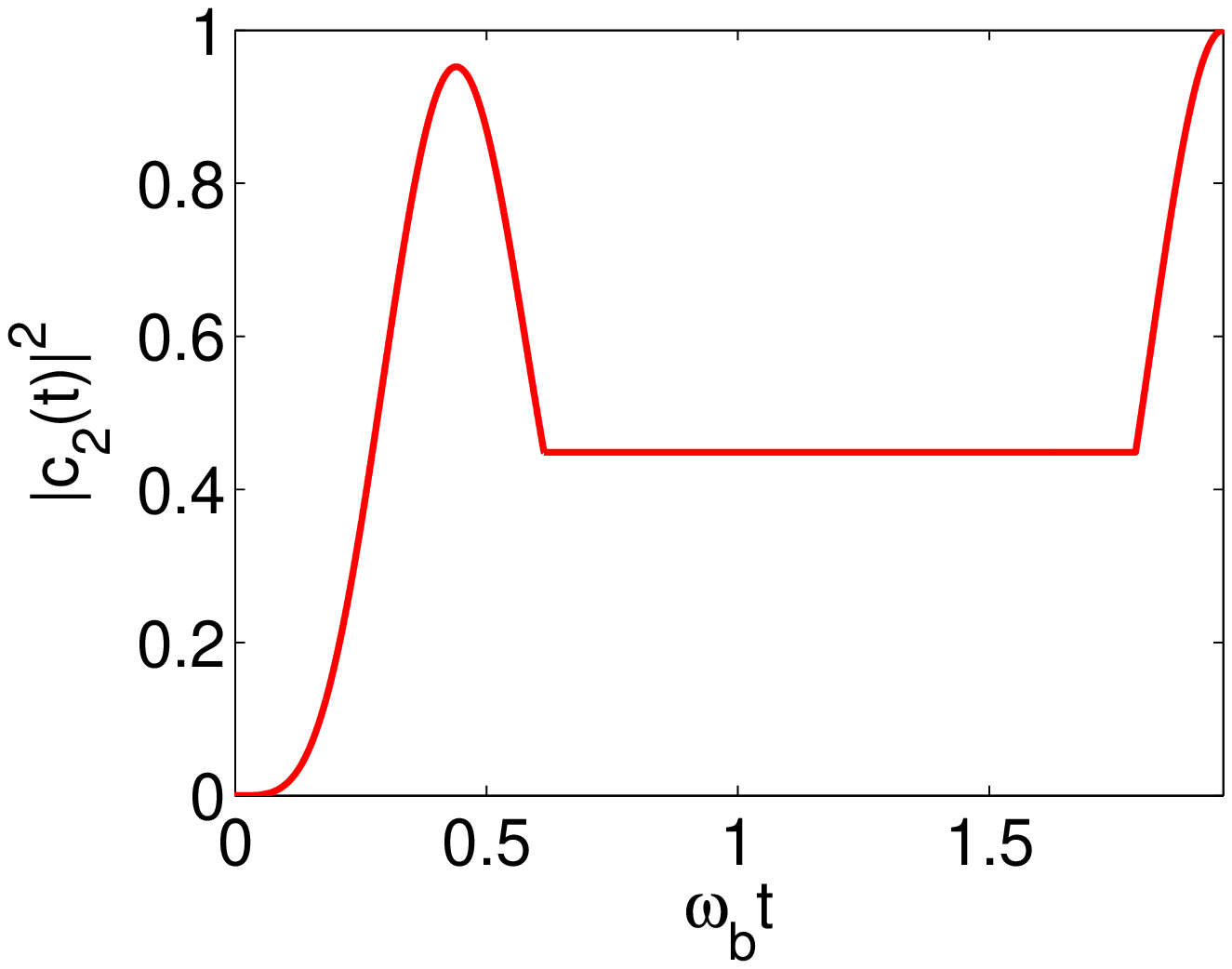}}
		\end{tabular}
\caption{(Color online) Two examples corresponding to maximum amplitudes $\Omega_0=4\sqrt{2/3}\omega_b$ (first row) and $\Omega_0=10\omega_b$ (second row). (a, b) Error (\ref{trans_error}) of the transcendental Eq. (\ref{transcendental}) versus the duration $\tau_2$ of the off segment. The negative resonances correspond to the solutions of this equations. Observe that for larger $\Omega_0$ there are more than one solutions, and we pick the largest one, highlighted with red star, leading to the shortest overall duration of the pulse-sequence. (c, d) Minimum-time on-off-on pulse-sequences. For the first example we display both the pulse-sequences, with interchanged $\tau_1, \tau_3$, which accomplish biexciton preparation within the same time. Note that before and after the state preparation the Rabi frequency is turned off. (e, f) Time evolution of population $|c_2(t)|^2$.}
\label{fig:example}
\end{figure*}

Observe from the two-level system initial and final conditions (\ref{i_f_conditions}) that the system should return to its initial state having acquired a $\pi$-phase. It is not hard to see that the simplest on-off pulse-sequence such that the system returns to the north pole of the Bloch sphere at the final time $t=T$ has the form on-off-on. The initial on-pulse (with the transverse field activated) removes the Bloch vector from the north pole. During the intermediate off-pulse the Bloch vector is rotated parallel to the equator, while the final on-pulse brings it back to the north pole. We now find the durations $\tau_i$, $i=1,2,3$, of these pulses such that the $\pi$-phase condition is satisfied in the minimum possible time $T=\tau_1+\tau_2+\tau_3$. For this pulse-sequence the propagator $U$ becomes
\begin{equation}
\label{propagator1}
U=U_{on}^{\tau_3}U_{off}^{\tau_2}U_{on}^{\tau_1},
\end{equation}
where $U_{on}^{\tau_j}$, $j=1,3$, is given by Eq. (\ref{U_on}) using the appropriate duration $\tau_j$, and $U_{off}^{\tau_2}$ corresponds to evolution under Hamiltonian $H_{off}=\omega_b\sigma_z$, obtained for $\Omega(t)=0$, i.e.
\begin{eqnarray}
\label{U_off}
U_{off}^{\tau_2}&=&e^{-iH_{off}\tau_2}=e^{-i\omega_b\tau_2\sigma_z}\nonumber\\
&=& I\cos{\omega_b\tau_2}-i\sin{\omega_b\tau_2}\sigma_z.
\end{eqnarray}
Using these expressions for the individual propagators and the following property of Pauli matrices
\begin{equation}
\label{Pauli}
\sigma_{\alpha}\sigma_{\beta}=\delta_{\alpha\beta}I+i\epsilon_{\alpha\beta\gamma}\sigma_\gamma,
\end{equation}
where $\alpha,\beta,\gamma$ can be any of $x,y,z$, $\delta_{\alpha\beta}$ is the Kronecker delta and $\epsilon_{\alpha\beta\gamma}$ is the Levi-Civita symbol,
we can express the full propagator $U$ as a linear combination of $\sigma_\alpha$ and the identity $I$,
\begin{eqnarray}
\label{propagator2}
U &=& u_II+u_x\sigma_x+u_y\sigma_y+u_z\sigma_z\nonumber\\
  &=&
\left(
\begin{array}{cc}
u_I+u_z & u_x+iu_y\\
u_x-iu_y & u_I-u_z
\end{array}
\right),
\end{eqnarray}
with coefficients
\begin{subequations}
\label{coefficients}
\begin{eqnarray}
\label{uI} u_I &=& \cos{\omega_b\tau_2}\cos{\omega(\tau_1+\tau_3)}\nonumber\\
               && -n_z\sin{\omega_b\tau_2}\sin{\omega(\tau_1+\tau_3)},\\
\label{ux} u_x &=& -in_x\cos{\omega_b\tau_2}\sin{\omega(\tau_1+\tau_3)}\nonumber\\
               && +2in_xn_z\sin{\omega\tau_1}\sin{\omega_b\tau_2}\sin{\omega\tau_3},\\
\label{uy} u_y &=& in_x\sin{\omega_b\tau_2}\sin{\omega(\tau_3-\tau_1)},\\
\label{uz} u_z &=& -in_z\cos{\omega_b\tau_2}\sin{\omega(\tau_1+\tau_3)}-i\sin{\omega_b\tau_2}\cos{\omega(\tau_3-\tau_1)}\nonumber\\
               && +2in_z^2\sin{\omega\tau_1}\sin{\omega_b\tau_2}\sin{\omega\tau_3}.
\end{eqnarray}
\end{subequations}

If we use Eq. (\ref{propagator2}) in condition (\ref{phase_condition}), from the off-diagonal elements we immediately obtain $u_x=u_y=0$. We start from $u_y=0$ and use Eq. (\ref{uy}). Note that the possible solution $\sin{\omega_b\tau_2}=0$ leads to $\tau_2=\pi/\omega_b$ and $T=\tau_1+\tau_2+\tau_3>\pi/\omega_b$, longer than the corresponding constant pulse solution, thus it is rejected. Setting equal to zero the other factor of Eq. (\ref{uy}), $\sin{\omega(\tau_3-\tau_1)}=0$, we obtain the symmetric relation $\tau_1=\tau_3$ and the asymmetric one
\begin{equation}
\label{asymmetric}
\tau_3-\tau_1=\pm\frac{\pi}{\omega}.
\end{equation}
We next show that the symmetric relation leads to a solution $T>\pi/\omega_b$ and thus is rejected. For $\tau_1=\tau_3$ Eq. (\ref{ux}) becomes
\begin{equation}
\label{u_x}
u_x=2in_x\sin{\omega\tau_1}(n_z\sin{\omega\tau_1}\sin{\omega_b\tau_2}-\cos{\omega\tau_1}\cos{\omega_b\tau_2}).
\end{equation}
For $u_x=0$, the solution $\sin{\omega\tau_1}=0$ leads to
\begin{equation}
\label{bad_U_1}
U=
\left(
\begin{array}{cc}
e^{-i\omega_b\tau_2} & 0\\
0 & e^{i\omega_b\tau_2}
\end{array}
\right),
\end{equation}
where note that we have also used Eq. (\ref{coefficients}) for $u_I, u_z$. Using this expression for $U$ in Eq. (\ref{phase_condition}) we get $e^{i\omega_bT}e^{-i\omega_b\tau_2}=-1\Rightarrow e^{i\omega_b(T-\tau_2)}=e^{i\pi}$, thus $T>\pi/\omega_b$ and the solution is rejected. The other solution of $u_x=0$,
\begin{equation}
\label{condition}
\cos{\omega\tau_1}\cos{\omega_b\tau_2}=n_z\sin{\omega\tau_1}\sin{\omega_b\tau_2},
\end{equation}
leads after some manipulation of the rest coefficients in Eq. (\ref{coefficients}) to
\begin{equation}
\label{bad_U_2}
U=
\left(
\begin{array}{cc}
-e^{i\omega_b\tau_2} & 0\\
0 & -e^{-i\omega_b\tau_2}
\end{array}
\right),
\end{equation}
thus $-e^{i\omega_bT}e^{i\omega_b\tau_2}=-1\Rightarrow e^{i\omega_b(T+\tau_2)}=e^{2i\pi}$ and $T+\tau_2=2\pi/\omega_b$. If we suppose that $T<\pi/\omega_b$ then from the last relation we get $\tau_2=2\pi/\omega_b-T>\pi/\omega_b$ and thus $T=\tau_1+\tau_2+\tau_3>\pi/\omega_b$, i.e. we arrive at a contradiction. The initial hypothesis is consequently wrong, thus $T\geq\pi/\omega_b$ and the corresponding solution is rejected.

We now move to study the non-symmetric case. Using Eq. (\ref{asymmetric}), we easily find for $u_x$ an expression similar to Eq. (\ref{u_x}), with a minus sign on the right hand side. The conditions for $u_x=0$ are the same as before. Solution $\sin{\omega\tau_1}=0$ leads to a propagator $U$ same as in Eq. (\ref{bad_U_2}) and thus to $T>\pi/\omega_b$, consequently is rejected. The other condition, given in Eq. (\ref{condition}), leads to the propagator
\begin{equation}
\label{good_U}
U=
\left(
\begin{array}{cc}
e^{i\omega_b\tau_2} & 0\\
0 & e^{-i\omega_b\tau_2}
\end{array}
\right).
\end{equation}
Note that in order to derive Eq. (\ref{good_U}), we have also used in Eq. (\ref{coefficients}) the trigonometric identities $\sin{(x\pm\pi)}=-\sin{x}, \cos{(x\pm\pi)}=-\cos{x}$.
Using Eq. (\ref{good_U}) in Eq. (\ref{phase_condition}) we obtain the condition $e^{i\omega_bT}e^{i\omega_b\tau_2}=-1\Rightarrow e^{i\omega_b(T+\tau_2)}=e^{i\pi}$, thus
\begin{equation}
\label{pi_phase}
T+\tau_2=\tau_1+2\tau_2+\tau_3=\frac{\pi}{\omega_b},
\end{equation}
which obviously leads to an acceptable solution (with duration shorter than the minimum time needed with a constant pulse $T<\pi/\omega_b$). Eqs. (\ref{asymmetric}), (\ref{condition}) and (\ref{pi_phase}) form a system for the unknown durations $\tau_i, i=1,2,3$. We can use Eqs. (\ref{asymmetric}), (\ref{pi_phase}) to eliminate $\tau_1, \tau_3$ and obtain
\begin{subequations}
\label{t13}
\begin{eqnarray}
\tau_1 &=& \frac{\pi}{2}\left (\frac{1}{\omega_b}\mp\frac{1}{\omega}\right)-\tau_2,\label{t1}\\
\tau_3 &=& \frac{\pi}{2}\left (\frac{1}{\omega_b}\pm\frac{1}{\omega}\right)-\tau_2.\label{t3}
\end{eqnarray}
\end{subequations}
Using Eq. (\ref{t1}) into Eq. (\ref{condition}) and the trigonometric identity $\tan{(x\pm\pi/2)}=-\cot{x}$, we end up with the following transcendental equation for middle duration $\tau_2$
\begin{equation}
\label{transcendental}
\tan\left[\omega(\frac{\pi}{2\omega_b}-\tau_2)\right] = -n_z\tan{\omega_b\tau_2}.
\end{equation}

From Eq. (\ref{t13}) it is obvious that, to each solution $\tau_2$ of Eq. (\ref{transcendental}) correspond two pairs of $\tau_1, \tau_3$ with interchanged values, i.e. if the pulse-sequence $(\tau_1=x, \tau_2, \tau_3=y)$ is a solution for biexciton preparation, then $(\tau_1=y, \tau_2, \tau_3=x)$ is also a solution with the same duration. Note that if the ratio $\Omega_0/\omega_b$ is lower than a certain threshold then Eq. (\ref{transcendental}) has no solution, i.e. a pulse sequence of the form on-off-on cannot prepare the biexciton state within a duration shorter than $\pi/\omega_b$. 
We can find this threshold by setting $\tau_2=0$ in Eq. (\ref{transcendental}). We obtain $\tan{[\pi\omega/(2\omega_b)]}=0$ and from this it turns out that the threshold value is $\Omega_0^{min}=\omega_b\sqrt{6}$, same as the amplitude of the constant pulse with minimum duration. For $\Omega_0<\Omega_0^{min}$, our experience from similar optimal control problems suggests that more than two switchings might be necessary in order to acquire the desired $\pi$-phase in a minimum time longer than $\pi/\omega_b$, see for example our recent work \cite{Stefanatos19,Stefanatos20}. Since values $\Omega_0>\Omega_0^{min}$ can be routinely achieved in the experiments and additionally lead to durations shorter than $\pi/\omega_b$, we concentrate in this range. Note that in this case Eq. (\ref{transcendental}) may have more than one solutions, as $\Omega_0$ increases. In such case we pick the solution with larger $\tau_2$ since it corresponds to a shorter duration $T=\pi/\omega_b-\tau_2$, as obtained from Eq. (\ref{pi_phase}). For very large values of $\Omega_0$ the shortest duration tends to the limiting value $\pi/(2\omega_b)$.

In Fig. \ref{fig:min_time} we plot with red solid line the duration $T$ corresponding to the minimum-time on-off-on pulse-sequence for values of the maximum amplitude in the range $\sqrt{6}\leq\Omega_0/\omega_b\leq 50$. Observe that for $\Omega_0=\omega_b\sqrt{6}$ this duration is $\pi/\omega_b$, while for large $\Omega_0$ tends to the limiting value $\pi/(2\omega_b)$. We demonstrate next that these minimum-time on-off-on solutions correspond actually to the numerically obtained minimum necessary time for the considered transfer. We use various values of $\Omega_0$ depicted in Fig. \ref{fig:min_time}, starting from $\Omega_0=4\sqrt{2/3}\omega_b$ (corresponding to the maximum amplitude of the hyperbolic secant pulse), then move to $\Omega_0=5\omega_b$, and use a step $\delta\Omega_0=5\omega_b$ afterwards. For each of these values we use numerical optimal control and specifically the BOCOP solver \cite{bocop}, to minimize the final error $P_e=1-|c_2(T)|^2$ when the control signal is restricted as $0\leq\Omega(t)\leq\Omega_0$, for various durations $T$ with a small step $\delta T=0.01/\omega_b$. For example, in Fig. \ref{fig:speed_lim_1} we plot the final error (in logarithmic scale) as a function of duration $T$ for $\Omega_0=4\sqrt{2/3}\omega_b$. Observe that, as the duration changes from $T=2.75/\omega_b$ to $T=2.76/\omega_b$, the error drops substantially, marking the minimum time needed for biexciton preparation when using the specific maximum amplitude. The corresponding duration $T=2.76/\omega_b$ is then plotted in Fig. \ref{fig:min_time} (cyan square). Analogously, in Figs. \ref{fig:speed_lim_2}, \ref{fig:speed_lim_3} we identify numerically the minimum times for $\Omega_0=10/\omega_b$ and $\Omega_0=50/\omega_b$, respectively. The corresponding durations are plotted in Fig. \ref{fig:min_time}, along with the minimum times for the rest representative values $\Omega_0$. Observe that, at least for the range of $\Omega_0$ that we explore here, these numerically found minimum times coincide with those obtained using the on-off-on pulse-sequence. We believe that this coincidence is not accidental, but strongly suggests that the presented pulse-sequence provides the speed limit for biexciton preparation when using system (\ref{H_trans}) with real control $\Omega(t)$, bounded as $0\leq\Omega(t)\leq\Omega_0$. The fact that the potential minimum-time trajectory includes only a few switchings is not surprising, if we recall that for the equivalent two-level system the transverse field is larger than the longitudinal. For example, in Ref. \cite{Boscain06} it is proved that in such case the minimum-time trajectory steering the north-pole to the south-pole on the Bloch sphere has only one switching.

Having demonstrated that the on-off-on pulse-sequence can prepare the biexciton state at the numerically obtained minimum time, we now present two specific examples. We start with the case where $\Omega_0=4\sqrt{2/3}\omega_b$, which is the maximum amplitude of the hyperbolic secant pulse previously used. In Fig. \ref{fig:trans3} we plot in logarithmic scale the error
\begin{equation}
\label{trans_error}
e=\left|\tan\left[\omega(\frac{\pi}{2\omega_b}-\tau_2)\right] + n_z\tan{\omega_b\tau_2}\right|^2
\end{equation}
between the two sides of transcendental equation (\ref{transcendental}), versus the duration $\tau_2$ of the off segment in the sequence. The negative resonance highlighted with a red star corresponds to the solution of this equation, $\tau_2=0.3861/\omega_b$, leading to a duration $T=\pi/\omega_b-\tau_2\approx 2.7555/\omega_b$. In Fig. \ref{fig:control3} we plot the two pulse-sequences corresponding to the values (\ref{t13}) of $\tau_1, \tau_3$, with green dashed-dotted line for the upper sign (shorter $\tau_1$) and blue dashed line for the lower sign (longer $\tau_1$). In Fig. \ref{fig:population3} we display the corresponding time-evolution of population $|c_2(t)|^2$. It is obviously much faster than the evolution under the hyperbolic secant pulse, shown in Fig. \ref{fig:sech_pop}, which requires a duration of about $4/\omega_b$ for biexciton preparation. As a second example we consider tha larger amplitude value $\Omega_0=10\omega_b$. In Fig. \ref{fig:trans10} we display the corresponding error (\ref{trans_error}), and observe that now there are more than one solutions of the transcendental equation (\ref{transcendental}). As we previously explained, we pick the largest solution $\tau_2=1.1763\omega_b$ highlighted with red star, leading to the shortest pulse-sequence duration $T=\pi/\omega_b-\tau_2\approx 1.9653/\omega_b$. In Fig. \ref{fig:control10} we plot the control $\Omega(t)$, only for the lower sign of Eq. (\ref{t13}) with longer $\tau_1$, while in Fig. \ref{fig:population10} the corresponding time evolution of population $|c_2(t)|^2$.

Closing this section, we would like to make some remarks regarding the suggested on-off pulse-sequences. The first observation is about the physics behind these control schemes. Since the control field $\Omega(t)$ couples both the ground-to-exciton and the exciton-to-biexciton transitions, the on-off modulation is used to create a destructive interference for the exciton state, so the population is transferred to the biexciton state. The second remark is about the duration of these pulse-sequences in the limit $\Omega_0\rightarrow\infty$, which is $\pi/(2\omega_b)$. From the previous analysis we tend to believe that this is the ultimate speed limit for perfect biexciton preparation, for the system under consideration and the accompanying control constraints.

\section{Effect of dissipation and dephasing}

\label{sec:relaxation}

\begin{figure*}[t]
 \centering
		\begin{tabular}{cc}
     	
      \subfigure[$\ $]{
	            \label{fig:dephase1}
	            \includegraphics[width=.45\linewidth]{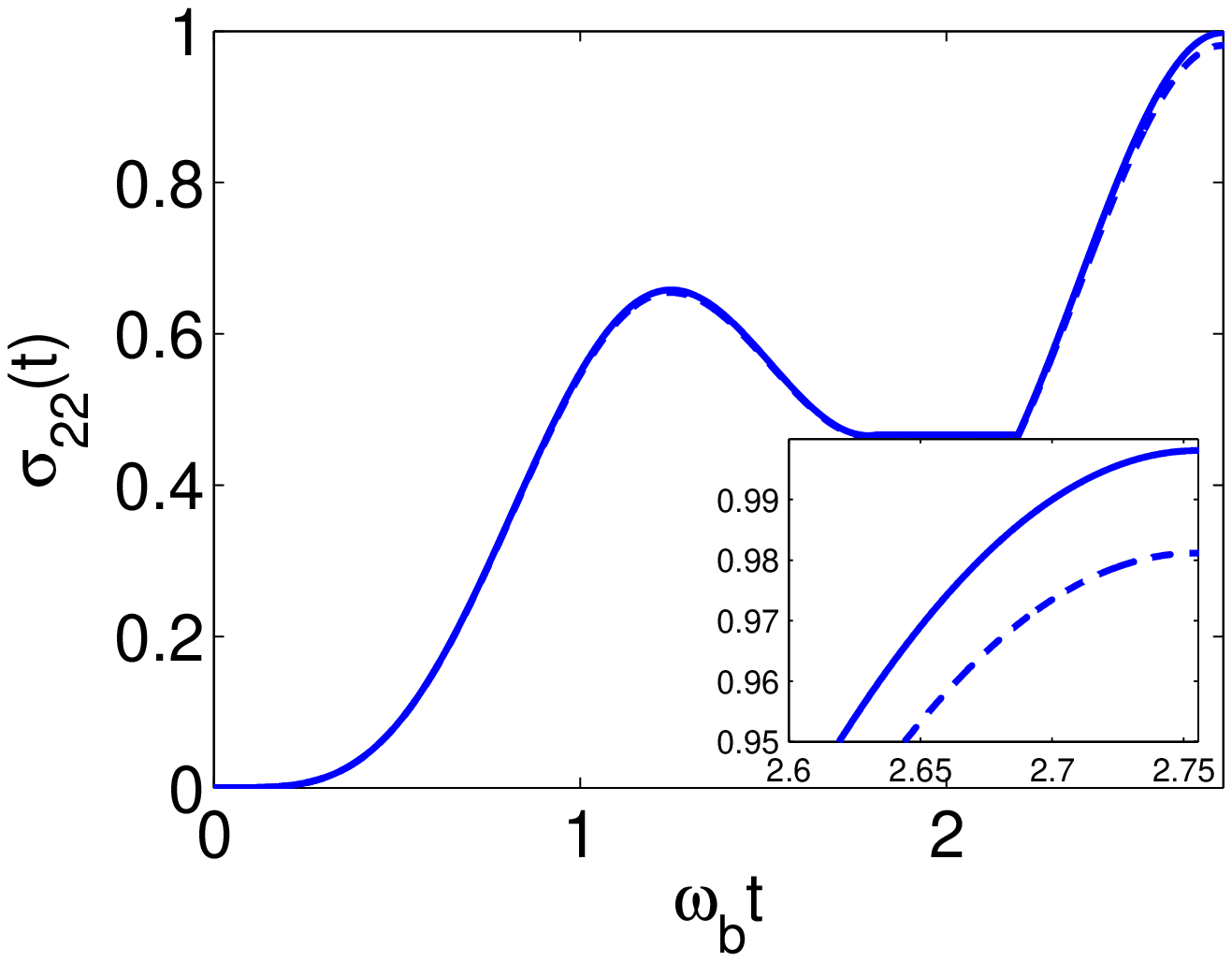}} &
      \subfigure[$\ $]{
	            \label{fig:dephase2}
	            \includegraphics[width=.45\linewidth]{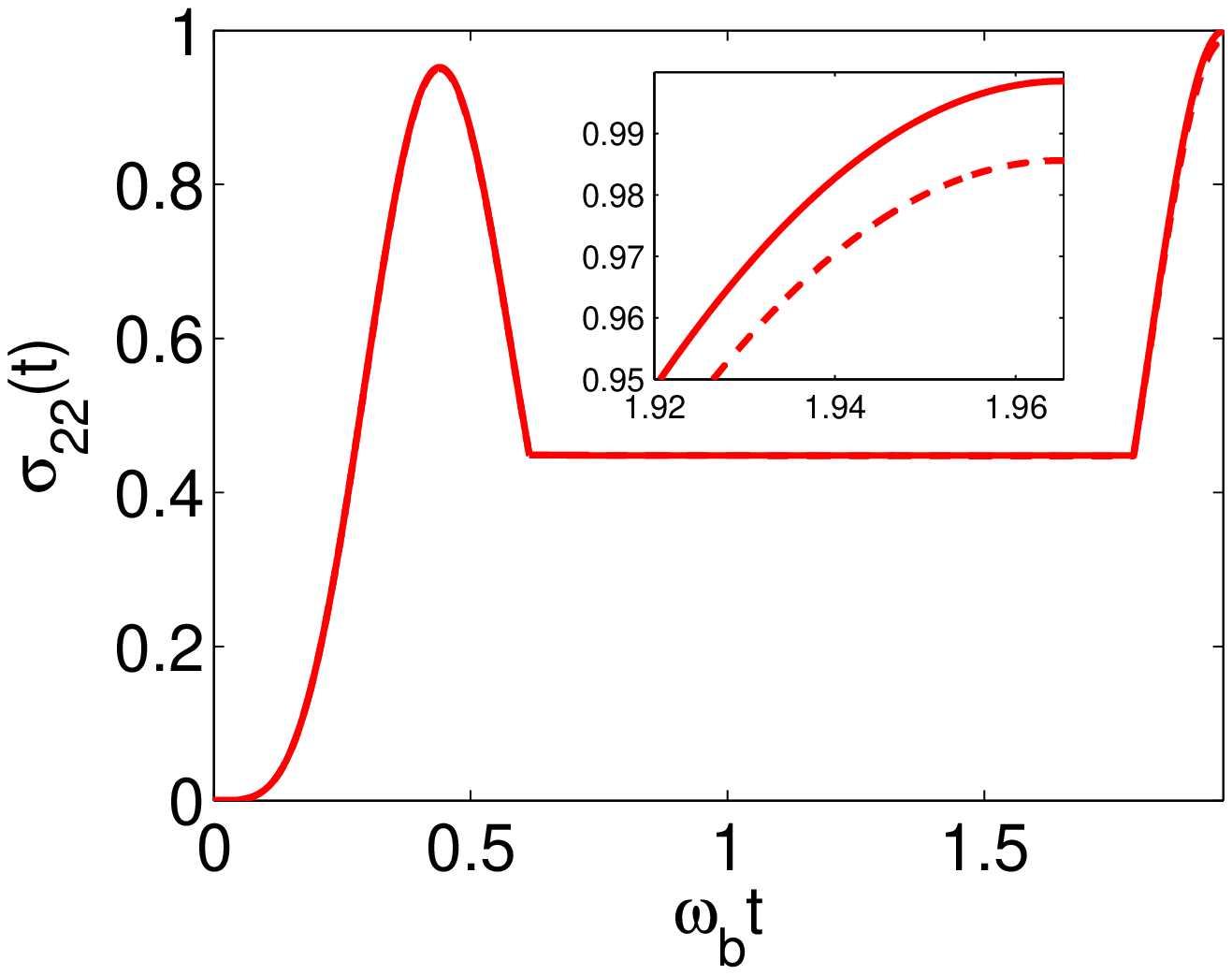}}
		\end{tabular}
\caption{(Color online) Time evolution of biexciton population $\sigma_{22}(t)$ in the presence of dissipation $\Gamma_{11}=\Gamma_{22}=\Gamma=1 \,\mbox{ns}^{-1}$ and dephasing $\gamma_{01}=\gamma_{02}=\gamma_{12}=\gamma=7\,\mbox{ns}^{-1}$, for two values of the biexciton frequency shift, $\omega_b=5\,\mbox{ps}^{-1}$ (solid lines) and $\omega_b=0.5\,\mbox{ps}^{-1}$ (dashed lines). (a) $\sigma_{22}(t)$ for the on-off-on pulse sequence with maximum amplitude $\Omega_0=4\sqrt{2/3}\omega_b$, displayed with blue dashed line in Fig. \ref{fig:control3}. (b) $\sigma_{22}(t)$ for the on-off-on pulse sequence with maximum amplitude $\Omega_0=10\omega_b$, shown in Fig. \ref{fig:control10}.}
\label{fig:speed_limit}
\end{figure*}

In order to incorporate in our study dissipation and dephasing, it is necessary to use the full density matrix equations of the system. Instead of using the density matrix elements, it is actually more convenient to use their slowly varying envelopes defined as $\rho_{nn}(t)=\sigma_{nn}(t)$, $n=1,2,3$, $\rho_{01}(t) = \sigma_{01}(t)e^{i \omega t}$, $\rho_{02}(t) = \sigma_{02}(t)e^{2 i \omega t}$, and $\rho_{12}(t) = \sigma_{12}(t)e^{i \omega t}$. Using Hamiltonian (\ref{ham1}), in the rotating wave approximation and at the two-photon resonance condition $\hbar\omega=E+E_b/2$, we find
\begin{subequations}
\label{pheno}
\begin{eqnarray}
\dot{\sigma}_{00}(t) &=& \Gamma_{11}\sigma_{11}(t)+i\frac{\Omega^{*}(t)}{2}\sigma_{10}(t) - i\frac{\Omega(t)}{2}\sigma_{01}(t), \\
\dot{\sigma}_{22}(t) &=& -\Gamma_{22}\sigma_{22}(t)+i\frac{\Omega(t)}{2}\sigma_{12}(t) - i\frac{\Omega^{*}(t)}{2}\sigma_{21}(t), \\
\dot{\sigma}_{01}(t) &=&-(2i\omega_b+\gamma_{01})\sigma_{01}(t)+i\frac{\Omega^{*}(t)}{2}\left[\sigma_{11}(t)-\sigma_{00}(t)\right]\nonumber\\
&&- i \frac{\Omega(t)}{2}\sigma_{02}(t), \\
\dot{\sigma}_{02}(t) &=&-\gamma_{02}\sigma_{02}(t)+i\frac{\Omega^{*}(t)}{2}\left[\sigma_{12}(t)-\sigma_{01}(t)\right], \\
\dot{\sigma}_{12}(t) &=&\left(2i\omega_b-\gamma_{12}\right)\sigma_{12}(t)+i\frac{\Omega^{*}(t)}{2}\left[\sigma_{22}(t)-\sigma_{11}(t)\right]\nonumber\\
&&+ i \frac{\Omega(t)}{2}\sigma_{02}(t),
\end{eqnarray}
\end{subequations}
where note that we have also incorporated the dissipation rates of the single-exciton and biexciton states, $\Gamma_{11}$ and $\Gamma_{22}$, and the dephasing rates of the off-diagonal matrix elements, $\gamma_{nm}, n\neq m$.

We consider specific examples using the values $\Gamma_{11}=\Gamma_{22}=\Gamma=1 \,\mbox{ns}^{-1}$, $\gamma_{01}=\gamma_{02}=\gamma_{12}=\gamma=7\,\mbox{ns}^{-1}$, taken from Ref. \cite{Paspalakis10a}. For the biexciton energy shift we will use the values $E_b/\hbar=20\,\mbox{ps}^{-1}$ and $E_b/\hbar=2\,\mbox{ps}^{-1}$, corresponding to $\omega_b=5\,\mbox{ps}^{-1}$ and $\omega_b=0.5\,\mbox{ps}^{-1}$, respectively. For the first value we have $\Gamma/\omega_b=2\times 10^{-4}, \gamma/\omega_b=1.4\times 10^{-3}$, while for the second $\Gamma/\omega_b=2\times 10^{-3}, \gamma/\omega_b=1.4\times 10^{-2}$. In Fig. \ref{fig:dephase1} we plot the time evolution of the biexciton population $\sigma_{22}(t)$ for the on-off-on pulse sequence with maximum amplitude $\Omega_0=4\sqrt{2/3}\omega_b$ (displayed with blue dashed line in Fig. \ref{fig:control3}), using blue solid line for $\omega_b=5\,\mbox{ps}^{-1}$ and blue dashed line for $\omega_b=0.5\,\mbox{ps}^{-1}$. From the plot and the detail shown in the inset becomes obvious that the protocol behaves quite well even in the presence of dissipation and dephasing, with the performance being better for larger $\omega_b$. The final population values for the two cases are $\sigma_{22}(T)=0.9981$ and $\sigma_{22}(T)=0.9812$. In Fig. \ref{fig:dephase2} we display similar plots for the pulse-sequence shown in Fig. \ref{fig:control10}, with maximum amplitude $\Omega_0=10\omega_b$. The exact values of the final population are now $\sigma_{22}(T)=0.9985$ and $\sigma_{22}(T)=0.9856$, for the larger and the smaller $\omega_b$, respectively. From this and the inset we conclude that with larger control amplitude a better efficiency is achieved within a shorter duration. Closing, we would like to point out that not all the important environmental interactions are captured by phenomenological rate equations accounting for dissipation and dephasing, like Eq. (\ref{pheno}). For example, we mention the phonon-assisted processes in optically driven quantum dots \cite{Axt14a}. Numerical optimal control \cite{bocop} can be used to maximize the biexciton state fidelity achieved within a fixed time interval, even in the presence of such processes.

\section{Conclusion}

\label{sec:conclusion}

We considered the problem of pulsed biexciton state preparation in a quantum dot and showed that, when using a on-off-on pulse-sequence with appropriate pulse durations, the target state can be reached faster than with the commonly used constant and hyperbolic secant pulses. Furthermore, we used numerical optimal control to show that the proposed pulse-sequence prepares the biexciton state in the numerically obtained minimum time for a wide range of the maximum pulse amplitude. Dissipation and dephasing have little effect on the system performance as long as the corresponding rates are small compared to the biexciton frequency shift. Since there is currently a significant demand for biexciton preparation schemes, e.g., for applications involving the generation of entangled photons \cite{Muller14a,Heinze15a,Winik17a,Huber17a,Chen18a}, we expect the present work to find application in this research area. Closing, it is worth mentioning that on-off pulses have also been suggested for preparation of maximally entangled states between superconducting gmon qubits \cite{Bao18}.

\begin{acknowledgments}
Co-financed by Greece and the European Union - European Regional Development Fund via the General Secretariat for Research and Technology bilateral Greek-Russian Science and Technology collaboration project on Quantum Technologies (project code name POLISIMULATOR).
\end{acknowledgments}

\end{document}